\documentclass[fleqn,10pt]{wlscirep}

\title{The effect of double counting, spin density, and Hund interaction in the different DFT+$U$ functionals}

\author[1]{Siheon Ryee}
\author[1,*]{Myung Joon Han}
\affil[1]{Department of Physics, KAIST, Daejeon 34141, Republic of Korea}

\affil[*]{mj.han@kaist.ac.kr}

\begin{abstract}
	A systematic comparative study has been performed to better understand DFT$+U$ (density functional theory + $U$) method. We examine the effect of choosing different double counting and exchange-correlation functionals. The calculated energy distribution and the Hund-$J$ dependence of potential profile for representative configurations clearly show the different behaviors of each DFT$+U$ formalism. In particular, adopting spin-dependent exchange-correlation functionals likely leads to undesirable magnetic solution. Our analyses are further highlighted by real material examples ranging from insulating oxides (MnO and NiO) to metallic magnetic systems (SrRuO$_3$ and BaFe$_2$As$_2$). The current work sheds new light on understanding DFT$+U$ and provides a guideline to use the related methods. 
\end{abstract}

\usepackage{graphicx}

\begin{document}
	
	\flushbottom
	\maketitle

	\thispagestyle{empty}

\section*{Introduction} \label{intro}

{\it Ab initio} description of strongly correlated materials has been a challenge in condensed matter physics and materials science. A promising and widely-used scheme is to combine local density approximation (LDA) with Hubbard-type model Hamiltonian approach within density functional theory (DFT) framework \cite{LDA++}. One of the earliest attempts of this kind is DFT+$U$ \cite{Anisimov_91,Liechtenstein,LDAU_review} which is now established as a standard approach. However, the calculation results of this type of methods strongly depends on the choice of double-counting energy functionals (which remove the conceptually equivalent contribution already present in LDA or GGA (generalized gradient approximation)) as well as interaction parameters (such as on-site Coulomb repulsion $U$ and Hund interaction $J$). This feature severely limits the predictive power of DFT$+U$ and its cousins such as DFT+DMFT (dynamical mean-field theory).

There have been many attempts to establish a proper double-counting scheme \cite{Anisimov_93,Czyzyk,Solovyev_94,Petukhov,nominal_dc,Amadon,Karolak,Wang,U',Haule}. 
The difficulty lies in the nonlinear dependence of exchange-correlation (XC) functionals on the charge and/or spin density. It is therefore non-trivial to extract the precise portion of LDA/GGA XC energy for the correlated subspace. Ever since its first invention of DFT+$U$ method, several phenomenological recipes have been suggested among which most widely used are so-called FLL (fully localized limit) \cite{Anisimov_93,Czyzyk,Solovyev_94,Liechtenstein} and AMF (around mean-field) \cite{Anisimov_91,Czyzyk}. Even though these double counting implementations have been extensively exploited, a comprehensive understanding of their working principles has not been reached. It is still unclear how and how much all these different formalisms give different results and predictions. In spite of previous analyses including some recent case studies of transition-metal systems within FLL \cite{Czyzyk,Solovyev_94,Petukhov,Bultmark,JChen,Park,Chen}, many functionals seem to be used often at random choice and without a proper guiding principle. As a result, it remains difficult to compare the results or predictions obtained by different DFT$+U$ formalisms.

In this paper, we perform a comparative study of representative DFT+$U$ functionals including FLL and AMF double countings. The effect of XC functional choice is also examined. To understand the detailed working principles of each DFT+$U$ formalism, we first examine the simplified model systems in terms of their energetics and potentials. Special attention has been paid to the $J$ dependence which has rarely been addressed before. Our analysis clearly shows the different behaviors of DFT+$U$ functionals and their origins. In particular, when spin-polarized version of LDA or GGA is adopted, it can likely produce the undesirable effects. The characteristic features are further highlighted with real material examples covering strongly correlated insulating oxides (MnO and NiO) and metallic magnetic systems (SrRuO$_3$ and BaFe$_2$As$_2$). Our work sheds new light on understanding DFT$+U$ formalism and related methodology, thereby providing an useful guideline for its applications.

\section*{Formalism}

In this section for the completeness and clarity of our presentation and notation, we briefly summarize DFT+$U$ formalisms within non-collinear density functional scheme. Simplification to collinear case is straightforward.
Hence `CDFT+$U$' refers to LDA+$U$ or GGA+$U$, and `SDFT+$U$' to LSDA+$U$ (local spin density approximation + $U$) or SGGA+$U$ (spin-polarized GGA + $U$). 
Also, we use terms ``cFLL''/``cAMF'' to denote CDFT+$U$ with FLL/AMF double counting and ``sFLL"/``sAMF'' to their SDFT+$U$ versions.

\subsection*{DFT+$U$ energy functionals}

DFT+$U$ total energy correction to CDFT or SDFT can be written as \cite{LDAU_review}:
\begin{eqnarray} 
{E^{U}} = \sum_{s}{E^{U}_s} = \sum_{s}{E^{\textrm {int}}_s}-E^{\textrm {dc}}_s,
\end{eqnarray} 
where $E^{\textrm{int}}_s$ and $E^\textrm{dc}_s$ refers to the  interaction energy within $d$- or $f$-shells and the double counting term, respectively for a particular atom $s$. From now on, we omit atom index $s$ for simplicity. In the present study, $E^U$ refers to either $E^U_{\textrm{FLL}}$ (FLL) or $E^U_{\textrm{AMF}}$ (AMF) depending on the choice of double counting term.

The FLL form of $E^\textrm{int}$ reads \cite{Liechtenstein,Fulde}:
\begin{align}  \label{int}
E^{\textrm{int}}_{\textrm{FLL}}  = \frac{1}{2}\sum_{\{m_i\},\sigma,\sigma'} \{n^{\sigma\sigma}_{m_1m_2}\langle m_1,m_3|V_{ee}|m_2,m_4 \rangle n^{\sigma'\sigma'}_{m_3m_4} - n^{\sigma\sigma'}_{m_1m_2} \langle m_1,m_3|V_{ee}|m_4,m_2 \rangle n^{\sigma'\sigma}_{m_3m_4} \},
\end{align}
where $n^{\sigma\sigma'}_{m_1m_2}$ are the elements of on-site density matrix (DM) $\mathbf{n}$ for orbitals $\{m_i\}$ and spins $\sigma,\sigma'$ ($\sigma,\sigma' = \uparrow$ or $\downarrow$)  \cite{MacDonald,Kubler}.
The matrix elements of on-site Coulomb interaction can be expressed by \cite{Liechtenstein, Vaugier}:
\begin{align} \label{Coulomb}
\langle m_1,m_3|V_{ee}|m_2,m_4 \rangle  = \sum_{\{m_i'\}}\Big[S_{m_1m_1'}S_{m_3m_3'} \Big\{\sum_{k=0}\alpha_k(m_1',m_3',m_2',m_4')F^k\Big\} S^{-1}_{m_2'm_2}S^{-1}_{m_4'm_4} \Big]
\end{align}
where $\alpha_k$ and $F^k$ refers to Racah-Wigner numbers and Slater integrals, respectively \cite{Liechtenstein,Vaugier}, and $S$ is a transformation matrix from spherical harmonics to the predefined local basis sets. 
We follow the conventional expression of $U=F^0$, $J=(F^2+F^4)/14$, and $F^4/F^2=0.625$ for $d$-orbitals. The effect of using different ratio between $F^4$ and $F^2$ is found to be negligible (see Supplementary Information).

Expressing $E^\textrm{dc}$ has long been an important issue and still remains as an open problem \cite{Karolak,Wang}. Note that $E^\textrm{dc}$ itself should depend on the given XC energy functional. The FLL double counting based on CDFT+$U$ (or cFLL) can be written as \cite{Anisimov_93,Solovyev_94}: 
\begin{align} \label{nFLL}
E_{\textrm {cFLL}}^{\textrm{dc}} = \frac{1}{2}UN(N-1) - \frac{1}{2}JN\bigg(\frac{N}{2}-1\bigg),
\end{align}
where $N=\textrm{Tr}[\mathbf{n}]$ within the correlated subspace. For SDFT+$U$ (or sFLL), effect of spin-polarized XC energy should also be taken into account \cite{Czyzyk,Liechtenstein,Bultmark}:
\begin{align} \label{FLL}
E_{\textrm {sFLL}}^{\textrm{dc}} = \frac{1}{2}UN(N-1) - \frac{1}{2}JN\bigg(\frac{N}{2}-1\bigg) -\frac{1}{4}J{\vec{\mathrm{M}} \cdot \vec{\mathrm{M}}},
\end{align} 
where the magentization $\vec{\mathrm{M}} = \textrm{Tr}[\vec{\sigma} \mathbf{n}]$ and $\vec{\sigma}$ is Pauli matrices \cite{Bultmark}. Note that the difference is the third term of Eq.~(\ref{FLL}). This formulation of Eq.~(\ref{FLL}) has been widely used.

In AMF formalism \cite{Anisimov_91,Czyzyk,Bultmark}, the energy correction is given by the fluctuation with respect to the average occupation of the correlated orbitals \cite{Anisimov_91}:
\begin{align} \label{int2}
{E^U_\textrm{AMF}} = E^{\textrm{int}}_{\textrm{AMF}} - E_\textrm{AMF}^\textrm{dc} = \frac{1}{2}\sum_{\{m_i\},\sigma,\sigma'} \{ \widetilde{n}^{\sigma\sigma}_{m_1m_2}\langle m_1,m_3|V_{ee}|m_2,m_4 \rangle 
\widetilde{n}^{\sigma'\sigma'}_{m_3m_4} - \widetilde{n}^{\sigma\sigma'}_{m_1m_2} \langle m_1,m_3|V_{ee}|m_4,m_2 \rangle \widetilde{n}^{\sigma'\sigma}_{m_3m_4} \},
\end{align}
where $\widetilde{n}^{\sigma\sigma'}_{m_1m_2}$ are the elements of the redefined DM $\mathbf{\widetilde{n}}$. In CDFT+$U$ (or cAMF) \cite{Anisimov_91},
\begin{align} \label{nAMF}
\mathbf{\widetilde{n}} = \mathbf{n} - \frac{1}{2(2l+1)}(N\mathbf{I}),
\end{align}
where $l$ denotes the angular momentum quantum number for the correlated subspace (e.g., $l = 2$ for $d$-shells) and $\mathbf{I}$ is the identity matrix. In SDFT+$U$ (or sAMF) \cite{Czyzyk,Bultmark},
\begin{align} \label{AMF}
\mathbf{\widetilde{n}} = \mathbf{n} - \frac{1}{2(2l+1)}(N\mathbf{I}+\vec{\sigma} \cdot \vec{\mathrm{M}}).
\end{align}

\subsection*{DFT+$U$ potentials}
The matrix elements of orbital dependent potentials are given by $V_{m_1m_2}^{U,\sigma \sigma'} = {\partial({E^{\textrm{int}}}-E^{\textrm{dc}}})/{\partial n^{\sigma \sigma'}_{m_1m_2}} = V_{m_1m_2}^{\textrm{int},\sigma \sigma'} - V_{m_1m_2}^{\textrm{dc},\sigma \sigma'}$. For FLL, the interaction potential for spin diagonal and off-diagonal part is given respectively by \cite{Liechtenstein,Fulde},
\begin{align}
\label{pot} &V_{\textrm{FLL},m_1m_2}^{\textrm{int},\sigma \sigma} = \sum_{m_3,m_4,\sigma'} \{\langle m_1,m_3|V_{ee}|m_2,m_4 \rangle  - \langle m_1,m_3|V_{ee}|m_4,m_2 \rangle \delta_{\sigma \sigma'} \}n^{\sigma' \sigma'}_{m_3m_4} 
\end{align}
and
\begin{align}
\label{Vint_off} V_{\textrm{FLL},m_1m_2}^{\textrm{int},\sigma \overline{\sigma}} &= - \sum_{m_3,m_4}\langle m_1,m_3|V_{ee}|m_4,m_2 \rangle n^{\overline{\sigma} \sigma}_{m_3m_4}.
\end{align}
Here, $\overline{\sigma}$ denotes the opposite spin to $\sigma$.
Within CDFT, the double counting potential is \cite{Anisimov_93,Solovyev_94}:
\begin{align} 
V_{\textrm{cFLL},m_1m_2}^{\textrm{dc},\sigma \sigma} &= \Big\{U\bigg(N-\frac{1}{2}\bigg) - J\bigg(\frac{N}{2}-\frac{1}{2}\bigg)\Big\}\delta_{m_1 m_2} 
\end{align}
and
\begin{align}
\label{VcFLL_off} V_{\textrm{cFLL},m_1m_2}^{\textrm{dc},\sigma \overline{\sigma}} = 0.\qquad \qquad \qquad \qquad \qquad \qquad \quad \;
\end{align}
Note that the off-diagonal components vanish and thus are spin independent. It is in a sharp contrast to the case of SDFT. In SDFT+$U$, \cite{Bultmark,Liechtenstein,Czyzyk}:
\begin{align}
&V_{\textrm{sFLL},m_1m_2}^{\textrm{dc},\sigma \sigma} = \Big\{U\bigg(N-\frac{1}{2}\bigg) - J\bigg(N^{\sigma \sigma}-\frac{1}{2}\bigg)\Big\}\delta_{m_1 m_2}, \\
\label{VsFLL_off} &V_{\textrm{sFLL},m_1m_2}^{\textrm{dc},\sigma \overline{\sigma}} = -JN^{\overline{\sigma} \sigma}\delta_{m_1 m_2},
\end{align}
where $N^{\sigma \sigma'}=\textrm{Tr}_m[\mathbf{n}^{\sigma \sigma'}]$ (taking trace over orbitals $m_i$). 

In AMF, the potential is given by taking derivative of Eq.~(\ref{int2}) with respect to density fluctuation $\mathbf{\widetilde n}$ \cite{Anisimov_91,Czyzyk,Bultmark}:
\begin{align}
V_{\textrm{AMF},m_1m_2}^{U,\sigma \sigma} &= \sum_{m_3,m_4,\sigma'} \{\langle m_1,m_3|V_{ee}|m_2,m_4 \rangle - \langle m_1,m_3|V_{ee}|m_4,m_2 \rangle \delta_{\sigma \sigma'} \}\widetilde{n}^{\sigma' \sigma'}_{m_3m_4}, \\
V_{\textrm{AMF},m_1m_2}^{U,\sigma \overline{\sigma}} &= - \sum_{m_3,m_4}\langle m_1,m_3|V_{ee}|m_4,m_2 \rangle \widetilde{n}^{\overline{\sigma} \sigma}_{m_3m_4},
\end{align}
where $\mathbf{\widetilde n}$ refers to Eq.~(\ref{nAMF}) and Eq.~(\ref{AMF}) for CDFT+$U$ (or cAMF) and SDFT+$U$ (or sAMF), respectively.

\section*{Analysis of model systems} \label{analysis}

To get a systematic understanding of how each DFT+$U$ functional works, we analyze model systems in this section. We investigate the behaviors of energy functionals and potentials as a function of key parameters, which provides useful insight into their differences.

\subsection*{Energetics} \label{energetics}
In general, DFT+$U$ DM is not necessarily diagonal \cite{Liechtenstein}. As it can always be diagonalized, however, we below assume the diagonalized DM without loss of generality.

Total energy corrections by DFT+$U$ in the case of collinear spins are now reduced to \cite{sum_rule,Czyzyk,Ylvisaker}:
\begin{align}
\label{cFLL} E^U_\textrm{cFLL} &= E^\textrm{int} - \frac{1}{2}UN(N-1) + \frac{1}{2}JN\bigg(\frac{N}{2}-1\bigg), \\
\label{sFLL} E^U_\textrm{sFLL} &= E^\textrm{int} - \frac{1}{2}UN(N-1) + \frac{1}{2}JN\bigg(\frac{N}{2}-1\bigg) + \frac{1}{4}JM^2, \\ 
\label{cAMF} E^U_\textrm{cAMF} &= E^\textrm{int} - \frac{1}{2}UN^2 + \frac{1}{4}\frac{U+2lJ}{2l+1}N^2, \\
\label{sAMF} E^U_\textrm{sAMF} &= E^\textrm{int} - \frac{1}{2}UN^2 + \frac{1}{4}\frac{U+2lJ}{2l+1}N^2 + \frac{1}{4}\frac{U+2lJ}{2l+1}M^2,
\end{align}
where
\begin{align}
E^\textrm{int} = \frac{1}{2}\sum_{\{m_i\},\sigma,\sigma'} n^{\sigma}_{m_1}\{\langle m_1,m_2|V_{ee}|m_1,m_2 \rangle - \langle m_1,m_2|V_{ee}|m_2,m_1 \rangle \delta_{\sigma\sigma'} \}  n^{\sigma'}_{m_2},
\end{align}
from  $n^{\sigma}_{m_1}=n^{\sigma \sigma'}_{m_1m_2}\delta_{m_1m_2}\delta_{\sigma \sigma'}$ in Eq.~(\ref{int}).
The fourth terms in Eq.~(\ref{sFLL}) and Eq.~(\ref{sAMF}) are responsible for the effective exchange interaction of SDFT (i.e., LSDA/SGGA, $U=0$). To represent the precise amount of this energy is a non-trivial task. Here we follow the conventional way of using Stoner parameter $I$ with which SDFT contribution to the energy gain via spin polarization is represented by $\Delta E^\textrm{SDFT} = -IM^2/4$ \cite{Andersen,Heine,Anisimov_91}.
Note that in sFLL, this contribution is cancelled out when
$J=I$ (see Eq.~(\ref{sFLL})). 

\begin{figure*} [!htbp]
	\includegraphics[width=0.99\textwidth, angle=0]{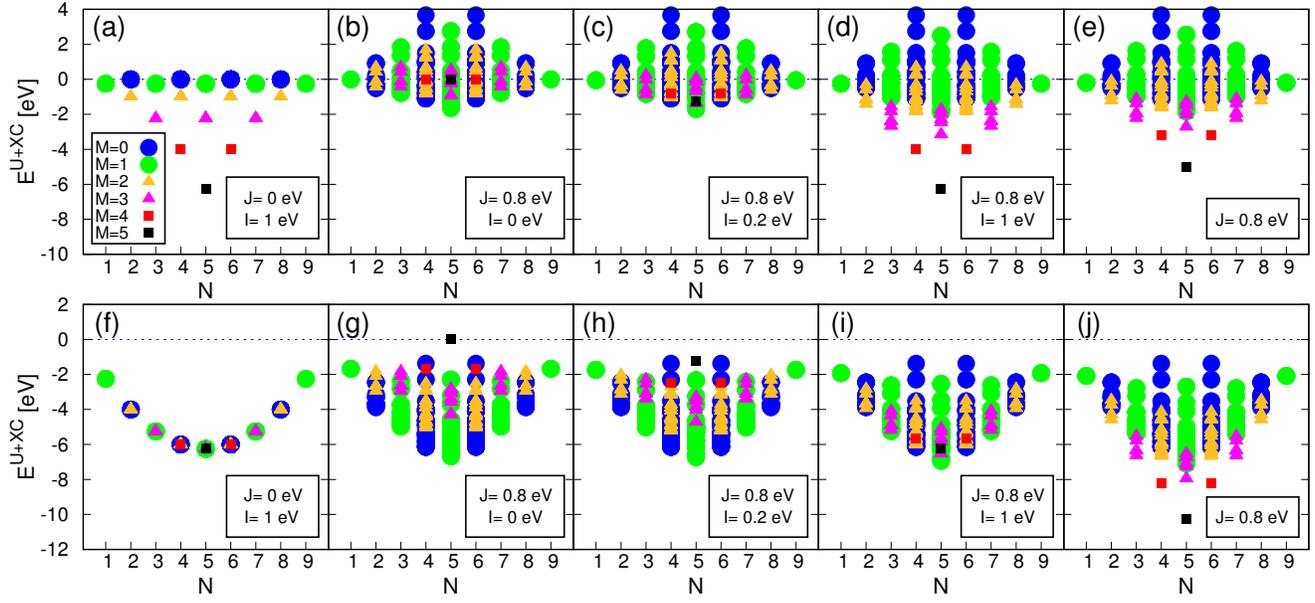}
	\caption{The energy distribution calculated by four different functionals; (a) - (d) sFLL, (e) cFLL, (f) - (i) sAMF, and (j) cAMF. $E^{U+\textrm{XC}}$ is defined as $E^{U+\textrm{XC}} = E^U_\textrm{sFLL(sAMF)} - IM^2/4$ for sFLL (sAMF) and $E^{U+\textrm{XC}} =  E^U_\textrm{cFLL(cAMF)}$ for cFLL (cAMF). All possible configurations with integer occupancy for given $N$ have been considered. The value of $U$ is fixed to 5 eV. }
	\label{energy}
\end{figure*}

Now let us see how these functionals work in different conditions. Before taking real material examples in the next section, we consider some idealized model systems. With a fixed value of $U=5$ eV, the energy distributions of $d$-shell electronic configurations are presented in Fig.~\ref{energy} (see also Fig.~3 of Ref.~\cite{Ylvisaker}). We use both $J$ and $I$ as control parameters.
Here, all possible configurations of integer occupancy for a given electron number $N$ are considered (e.g., $_{10}C_4=210$ configurations for $N=4$).
We present the energy from DFT+$U$ and XC functional contributions, which is defined as $E^{U+\textrm{XC}} \equiv E^U_\textrm{sFLL(sAMF)} - IM^2/4$ for sFLL (sAMF) and $E^{U+\textrm{XC}} \equiv E^U_\textrm{cFLL(cAMF)}$ for cFLL (cAMF).

Fig.~\ref{energy}(a) shows the result of $J=0$ which can represent so-called `simplified rotationally invariant' formalism by Dudarev {\it et al.} \cite{Dudarev}. Note that the configurations with the same $N$ are degenerate within $E^\textrm{sFLL}$ and this degeneracy is lifted by SDFT energy of $-IM^2/4$. Therefore, the largest possible $M$ configuration is always favored energetically. By comparing Fig.~\ref{energy}(a) with (d), one can clearly notice the role of $J$; lifting degeneracy within the same $N$-$M$ configurations \cite{Ylvisaker}. 

If the energy contribution from SDFT is negligible (i.e., $I=0$ in $\Delta E^\textrm{SDFT}$; Fig.~\ref{energy}(b)), the smaller $M$ configurations are favored. Only when it becomes significant (Fig.~\ref{energy}(d)), the larger $M$ states are stabilized and the Hund's first rule is satisfied. While sFLL has been considered to be appropriate for high spin systems \cite{Ylvisaker}, this behavior is mainly attributed to SDFT exchange rather than to DFT+$U$ correction, $E^{U}_\textrm{sFLL}$, as clearly seen by comparing Fig.~\ref{energy}(b) and (d). 
In sFLL, the low spin or nonmagnetic solution is favored as far as $J$ is significantly larger than $I$; see Fig.~\ref{energy}(c).

In cFLL, the spin state is controlled solely by the term $E^U_\textrm{cFLL}$. Note that Fig.~\ref{energy}(e) is quite similar with Fig.~\ref{energy}(d). If $I=J$ in sFLL, the third term in Eq.~(\ref{sFLL}) cancels $\Delta E^\textrm{SDFT}$ contribution and sFLL becomes equivalent to cFLL. If the exchange contribution implicit in SDFT is larger than $J$ (i.e., $I>J$), sFLL favors the larger $M$ state more than cFLL (compare Fig.~\ref{energy}(d) and (e)).

The estimation of the intrinsic exchange in SDFT is not trivial and in general material dependent. Recent works reported that it is about $\sim 1.0 - 1.5$ eV for 3$d$ transition metal systems such as nickelates, SrMnO$_3$, SrVO$_3$, and bcc Fe, which can be regarded as large \cite{Park,Chen}. As shown in Fig.~\ref{energy}(b) - (d), the exchange contribution from SDFT plays a major role in determining the moment formation, and therefore sFLL can prefer the unphysically large moment solutions. Further, SGGA has in general the stronger tendency toward the magnetic solution than LSDA \cite{Ryee}, which is another source of ambiguity. It is certainly a drawback of SDFT+$U$ especially for predicting material property.

In the case of AMF, the difference between CDFT and SDFT is more dramatic; see Fig.~\ref{energy}(f)--(j). As studied by Ylvisaker {\it et al.} \cite{Ylvisaker}, sAMF favors the low spin state and requires quite large value of $I$ to recover Hund's first rule. As shown in Fig.~\ref{energy}(i), sAMF still favors the lowest moment solution even for $I=1$ eV, which is in a sharp contrast to cAMF favoring the moment formation as in cFLL (Fig.~\ref{energy}(e) and (j)). It is attributed to the fourth term of Eq.~(\ref{sAMF}) which penalizes the larger moment formation. For example, with $U=5$ eV, $\frac{1}{4}\frac{U+2lJ}{2l+1}M^2 = \frac{1}{4}(1+\frac{4}{5}J)M^2$. Thus $I$ should be greater than $1+4J/5$ for exchange energy gain by SDFT. This feature can cause some practical problems in using AMF functionals.

\subsection*{$J$-dependence of potentials} \label{splitting_analysis}

To understand the effect of $J$ on the moment formation and spectral property, here we further analyze DFT+$U$ potentials. The $J$-only contribution to DFT+$U$ potentials (separated from $U$ contributions) for an orbital $m$ and spin $\sigma$ can be expressed as (assuming the diagonalized DM):
\begin{align} 
\label{VcFLL} \widetilde{V}^{U,\sigma}_{\textrm{cFLL},m} &= \widetilde{V}^{\textrm{int},\sigma}_{J,m} + J\bigg(\frac{N}{2}-\frac{1}{2}\bigg),\\
\label{VsFLL} \widetilde{V}^{U,\sigma}_{\textrm{sFLL},m} &= \widetilde{V}^{\textrm{int},\sigma}_{J,m} + J\bigg(N^\sigma - \frac{1}{2}\bigg), \\
\label{VcAMF} \widetilde{V}^{U,\sigma}_{\textrm{cAMF},m} &= \widetilde{V}^{\textrm{int},\sigma}_{J,m} + J\bigg(\frac{2l}{2l+1}\frac{N}{2}\bigg),\\
\label{VsAMF} \widetilde{V}^{U,\sigma}_{\textrm{sAMF},m} &= \widetilde{V}^{\textrm{int},\sigma}_{J,m} + J\bigg(\frac{2l}{2l+1}N^{\sigma}\bigg),
\end{align}
where $\widetilde{V}^{\textrm{int},\sigma}_{J,m}$ is obtained from Eq.~(\ref{pot}) by taking non-monopole terms in Coulomb interaction matrix elements,
\begin{align} \label{Vint2}
\widetilde{V}_{J,m_1}^{\textrm{int},\sigma} = \sum_{m_2,\sigma'} \{\langle m_1,m_2|V_{J,ee}|m_1,m_2 \rangle 
 - \langle m_1,m_2|V_{J,ee}|m_2,m_1 \rangle \delta_{\sigma \sigma'} \}n^{\sigma'}_{m_2},
\end{align}
and $\langle m_1,m_2|V_{J,ee}|m_1,m_2 \rangle$ is defined as
\begin{align} 
\langle m_1,m_2|V_{J,ee}|m_1,m_2 \rangle = \sum_{\{m_i'\}}\Big[S_{m_1 m_1'}S_{m_2 m_3'} \Big\{\sum_{k\ne 0}\alpha_k(m_1',m_3',m_2',m_4')F^k\Big\} S^{-1}_{m_2' m_1}S^{-1}_{m_4' m_2} \Big].
\end{align}
In Eq.~(\ref{VcFLL}) - (\ref{VsAMF}), the second terms are double counting contributions. 

One can clearly notice that sFLL and sAMF potentials have the spin-dependent double counting which causes the additional up/down spin potential difference. Namely, the spin-splitting is affected by double counting terms. For $\widetilde{V}^{U,\sigma}_{\textrm{cFLL},m}$ and $\widetilde{V}^{U,\sigma}_{\textrm{cAMF},m}$, on the other hand, the spin-splitting is only controlled by interaction potential, $\widetilde{V}^{\textrm{int},\sigma}_{J,m}$.

In Fig.~\ref{potential}, the calculated $J$-induced spin-splittings for the model systems are presented (see also Table~\ref{conf} for the list of configurations) . The potential difference, $\Delta \widetilde{V}^U_{\alpha} \equiv \widetilde{V}^{U,\downarrow}_{\alpha} - \widetilde{V}^{U,\uparrow}_{\alpha}$ for a given orbital $\alpha$, can be estimated in the unit of $J$ through Eq.~(\ref{VcFLL}) - (\ref{VsAMF}) and Eq.~(\ref{Vint2}).
Noticeable is the same behavior of cFLL and cAMF, in which $\Delta \widetilde{V}^U_{\alpha}$ is quite substantial and always positive, favoring the moment ($M$) formation. This feature is attributed to the spin potential in Eq.~(\ref{VcFLL}) and (\ref{VcAMF}) where the spin-splitting is only controlled by $\widetilde{V}^{\textrm{int},\sigma}_{J,m}$ due to the exact cancellation of up- and down-spin double counting potentials. Thus, it is not specific to a particular form of double counting scheme. Note that the effect of $J$ in CDFT+U (cFLL and cAMF) is consistent with what is expected from Hartree-Fock approximation. 

Very different features are found in sFLL where the sign of $\Delta \widetilde{V}^U_{\alpha}$ depends on the configuration. In particular, for configurations of $M \ge 3$ (i.e., configuration 8 -- 12), sFLL suppresses the spin-splittings, which is the case of SrMnO$_3$ reported by Chen {\it et al.} \cite{Chen} (see configuration 8). The trend of suppressing spin-splitting is most pronounced at half-filling (configuration 12), e.g., MnO. Further, it is important to note that the negative spin-splitting is not a general feature of sFLL double counting contrary to what is speculated by Ref.~\cite{Chen}. See the positive $\Delta \widetilde{V}^U_{\alpha}$ configurations in Fig.~\ref{potential}. Our result clearly shows that both sFLL and sAMF can produce the positive spin-splitting potential.

We note that SDFT+$U$ (sFLL and sAMF) behaves in a counter-intuitive way from the point of view of Hartree-Fock picture. It is because the spin-dependent double countings do not in general cancel out the exchange interaction from SDFT. To recover the Hartree-Fock behavior, it is desirable to use CDFT+$U$.

\section*{Application to real materials} \label{real}

\subsection*{Calculation detail} \label{detail}
All calculations were performed using our new implementation of DFT+$U$ into OpenMX software package \cite{openmx}, which is based on the nonorthogonal LCPAO (linear combination of localized pseudoatomic orbitals) formalism \cite{LCPAO1,LCPAO2,LCPAO3}. We adopted Troullier-Martins type norm-conserving pseudopotentials \cite{TM} with partial core correction. We used 9 $\times$ 9 $\times$ 9, 12 $\times$ 12 $\times$ 12 (8 $\times$ 8 $\times$ 6), and 14 $\times$ 14 $\times$ 7 $\mathbf{k}$-points for rocksalt MnO and NiO, cubic (orthorhombic {\it Pbnm}) SrRuO$_3$, and BaFe$_2$As$_2$ in the first Brillouin zone, respectively, and the energy cutoff of 500 Ry for numerical integrations in real space grid. The localized orbitals were generated with radial cutoff of 6.0 (Mn, Ni, and Fe) and 7.0 (Ru) a.u. \cite{LCPAO1,LCPAO2}. Experimental lattice parameters were used for all materials. For the XC functional, L(S)DA \cite{CA} parameterized by Perdew and Zunger \cite{CA-PZ} was used. Unless otherwise specified, we adopted `dual' projector \cite{MJH} for on-site DM. For more discussion on local projectors in LCPAO scheme, see Ref.~\cite{MJH}. 

\subsection*{MnO and NiO} \label{mno_and_nio}

\begin{figure} [!htbp]
	\includegraphics[width=0.7\textwidth, angle=0]{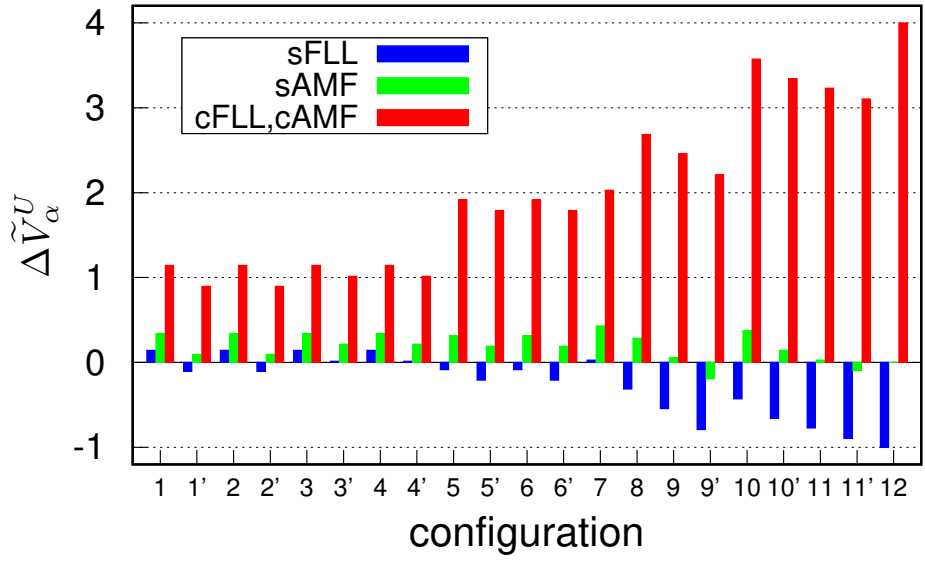}
	\caption{The calculated $J$-induced spin splitting $\Delta \widetilde{V}^U_{\alpha} \equiv \widetilde{V}^{U,\downarrow}_{\alpha} - \widetilde{V}^{U,\uparrow}_{\alpha}$ for the configurations and a given orbital $\alpha$ defined in Table~\ref{conf}. The results are presented in the unit of $J$.}
	\label{potential}
\end{figure}

Now we consider real materials. The first examples are MnO and NiO, corresponding to the configuration 12 and 7 in Fig.~\ref{potential}, respectively (see also Table~\ref{conf}). Although these two prototype correlated insulators have been extensively studied by using DFT+$U$, the systematic $J$-dependence of the electronic and magnetic property has rarely been addressed.

\begin{figure} [!htbp]
	\includegraphics[width=0.7\textwidth, angle=0]{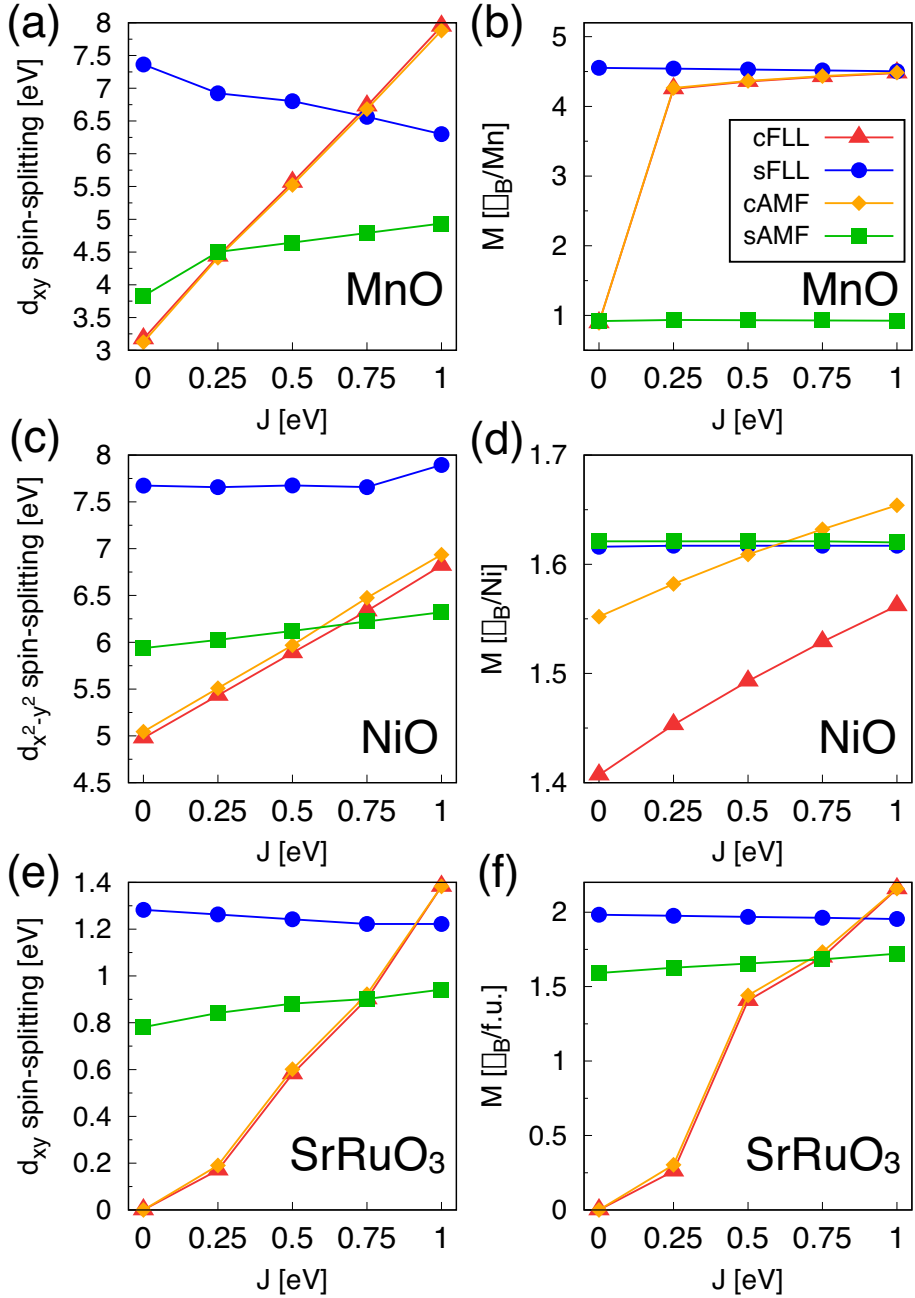}
	\caption{The $J$ dependence of spin-splitting ($\Delta \widetilde{V}^U_{\alpha}$) and $M$ in the ground states of (a, b) MnO, (c, d) NiO, and (e, f) cubic SrRuO$_3$. $U=$ 3, 5, and 2 eV for MnO, NiO and cubic SrRuO$_3$, respectively. Left and Right panels present the spin-splitting and $M$, respectively. In MnO and NiO, the energy level corresponding to orbital $\alpha$ and spin $\sigma$ is quantified by its center of mass position of DOS; $E^\sigma_{\alpha}= \int{ Eg^{\sigma}_{\alpha}(E)dE}/\int{g^{\sigma}_{\alpha}(E)dE}$, where $g^\sigma_\alpha(E)$ is DOS for given $\alpha$ and $\sigma$ at the energy $E$. For SrRuO$_3$, due to the strong Ru $d$ - O $p$ hybridization, the spin-splitting was estimated by the up- and down-spin difference of DOS peak position.}
	\label{split}
\end{figure}

In Fig.~\ref{split}, the calculated spin-splittings and magnetic moments by four different DFT+$U$ formalisms (namely, cFLL, sFLL, cAMF, and sAMF) are compared as a function of $J$. First of all, we note that the calculated $\Delta \widetilde{V}^U_{\alpha}$ is consistent with our analyses presented in Fig.~\ref{potential}. In MnO, the splitting is rapidly increased in cFLL and cAMF as $J$ increases, which is consistent with the positive value of $\Delta \widetilde{V}^U_\alpha$ in Fig.~\ref{potential}. On the other hand, it is gradually reduced in sFLL as a function of $J$, being consistent with the small and negative $\Delta \widetilde{V}^U_\alpha$ in Fig.~\ref{potential}. The results of NiO are also very well compared with the configuration 7 in Fig.~\ref{potential}

It is noted that sAMF predicts the entirely wrong magnetic ground state, $M \simeq 1$ $\mu_B$/Mn (see green lines in Fig.~\ref{split}(a) and (b)). This low spin configuration is no longer represented by configuration 12 in Fig.~\ref{potential}. This is an outstanding example to show that sAMF can unphysically favor the low spin state due to the overestimated $I$. In this kind of case, the use of sAMF is highly undesirable.

\begin{figure} [!htbp]
	\includegraphics[width=0.7\textwidth, angle=0]{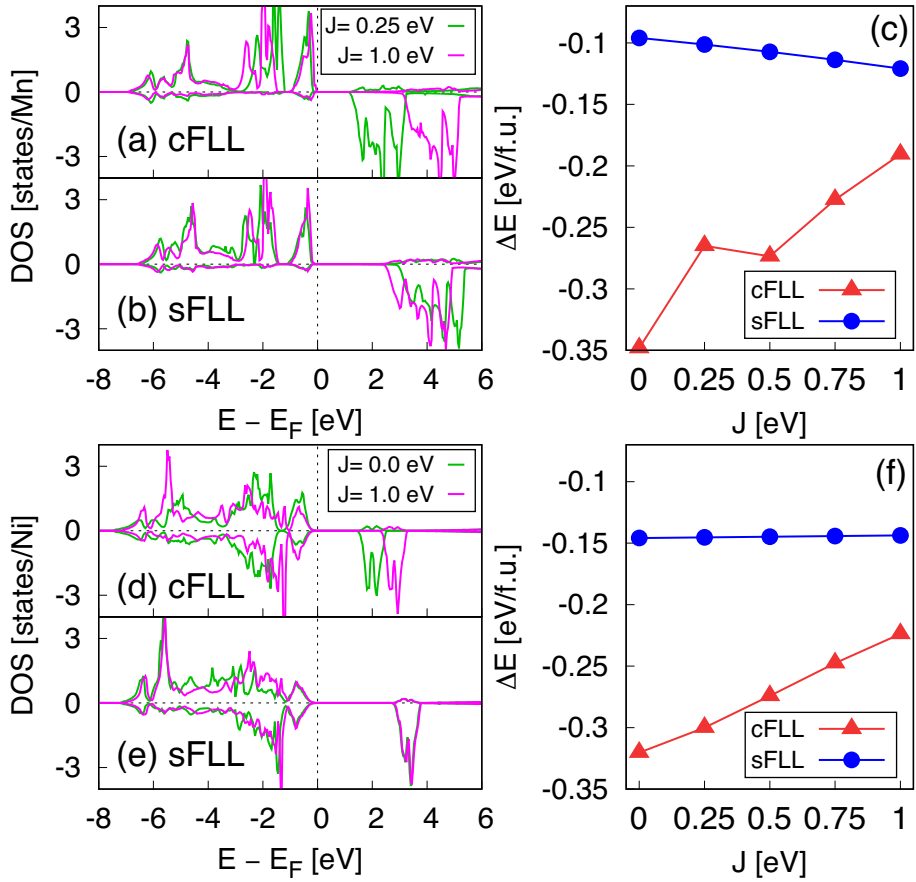}
	\caption{(a-c) The calculated Mn $d$ DOS by (a) cFLL and (b) sFLL within high-spin configurations. (c) The total energy difference $\Delta E$ for MnO as a function of $J$. (d-f) The calculated Ni $d$ DOS by (d) cFLL and (e) sFLL. (f) $\Delta E$ for NiO. The upper and lower panels in the DOS plots represent up and down spin parts, respectively.}
	\label{mno_nio}
\end{figure}

The high spin ground state of MnO is well reproduced by sFLL, cFLL, and cAMF in a reasonable range of $J$ (Fig.~\ref{split}(a) and (b)). In sFLL, this ground state configuration is obtained even at $J=0$ eV due to the intrinsic exchange within SDFT ($U=0$) large enough to stabilize the high spin. The calculated density of states (DOS) in Fig.~\ref{mno_nio}(a) and (b) clearly shows the different $J$ dependence of cFLL and sFLL functionals. While the up/down spin state split is mainly controlled by $J$ in cFLL, it is quite significant already at small $J$ in the case of sFLL.

To further elucidate the difference between CDFT+$U$ and SDFT+$U$, Fig.~\ref{mno_nio}(c) shows the total energy difference between antiferro- and ferro-magnetic phases ($\Delta E = E_\textrm{AF}-E_\textrm{FM}$) calculated by cFLL and sFLL. The $J$ dependence of $\Delta E$ exhibits the opposite trends; as $J$ increases, cFLL tends to less favor the AF order while sFLL more favors it. From the superexchange magnetic coupling of $J_\textrm{ex} \sim -t^2/(U+4J)$ ($t$: Mn-site effective hopping integral), the behavior predicted by cFLL is more reasonable than sFLL.

In NiO (Fig.~\ref{split}(c) and (d)), the $M$ is insensitive to $J$, $M \simeq 1.6$ $\mu_B$/Ni while the slight increase is observed in cFLL and cAMF following the trend of the $d_{x^2-y^2}$ spin-splitting (see also Fig.~\ref{mno_nio}(d) and (e)). Here we note that in this $d^8$ case the low and high spin configuration is irrelevant to get the ground state property. The calculated $\Delta E$ change is also quite small in sFLL (Fig.~\ref{mno_nio}(f)). In cFLL, $\Delta E = -0.320$ and $-0.224$ eV/f.u. at $J=0$ and $1$ eV, respectively, being consistent with superexchange estimation.

\subsection*{SrRuO$_3$}

SrRuO$_3$ is a ferromagnetic metal with a transition temperature of $T_c \sim 160$ K \cite{Koster}.
DFT+$U$ has often been used to study SrRuO$_3$ \cite{Jeng,Mahadevan,Granas,Verissimo} in spite of its metallic nature \cite{Georges_Hund}. Therefore it will be informative to investigate the DFT+$U$ functional dependence in this material. The configuration 6' in Fig.~\ref{potential} and Table~\ref{conf} corresponds to this case. Fig.~\ref{split}(e) and (f) shows the calculated spin-splitting and magnetic moment, respectively. They are consistent with the results of Fig.~\ref{potential}; namely, the slight decreasing (increasing) trend of splitting and moment in sFLL (sAMF) and the large increase in cFLL and cAMF as a function of $J$.

It is noted that sFLL gives the fully polarized spin moment of $M \simeq 2$ $\mu_B$/f.u. for both cubic and distorted orthorhombic (not shown) structures. This half-metallic phase has been reported before by using sFLL version of SDFT+$U$ \cite{Jeng,Mahadevan,Granas}, however, it is not well supported by experiments. The result of sAMF shows the smaller spin splitting and moment than those of sFLL as also reported in Ref.~\cite{Granas}. This behavior of sAMF and sFLL are consistent with what is observed in MnO and NiO discussed above. Namely, it is attributed to the spin-dependent double counting which depends on $U$ as well as $J$ in sAMF (Eq.~(\ref{sAMF})). Due to its metallic nature, the magnetism of SrRuO$_3$ can be more sensitive to the choice of double counting.

CDFT+$U$ (i.e., cFLL and cAMF) shows notably different behaviors. The calculated magnetic moment and splitting are gradually increased as a function of $J$ (Fig.~\ref{split}(e) and (f)) and the half-metallic phase is observed only for large $J$ ($J \gtrsim 0.9$ eV for cubic and $0.8$ eV for orthorhombic structure). In a reasonable range of $J \simeq 0.4$ -- $0.6$ eV \cite{comment,Si,Dang}, the calculated moment is $M \simeq 1.4$ and $1.6$ $\mu_B$/f.u. for cubic and orthorhombic structure, respectively, in good agreement with experiments \cite{Koster}.

\begin{figure} [!htbp]
	\includegraphics[width=0.9\textwidth, angle=0]{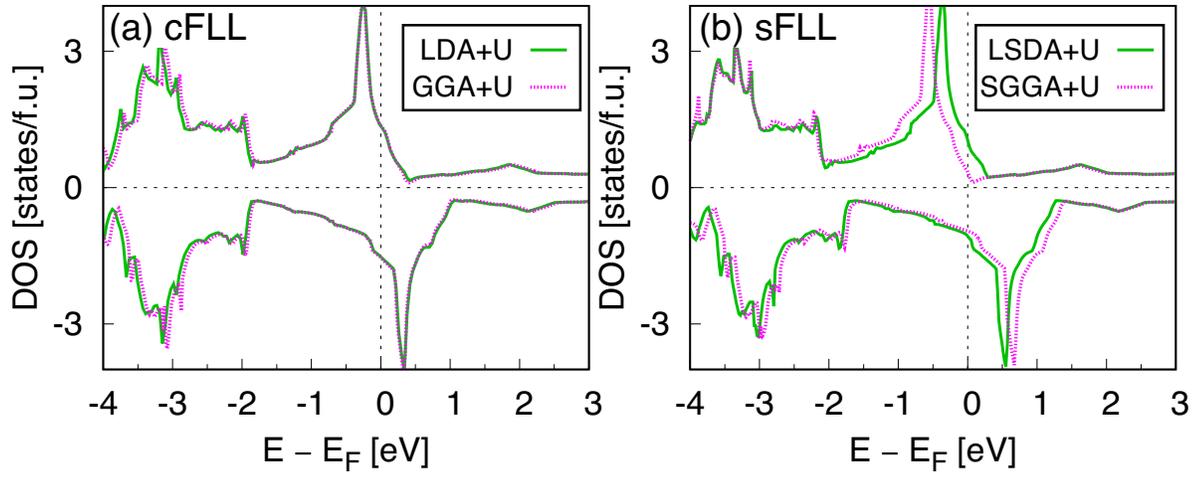}
	\caption{The calculated DOS of cubic SrRuO$_3$ by (a) cFLL and (b) sFLL. Two XC functionals for CDFT and SDFT are adopted; L(S)DA (green solid lines) and (S)GGA (magenta dotted lines). PBE (Perdew-Burke-Ernzerhof) parameterization for (S)GGA \cite{PBE} is used. $U=2$ and $J=0.5$ eV for cFLL (a), and $U=1$ and $J=0$ eV for sFLL (b).}
	\label{sro_dos}
\end{figure}

As mentioned in the previous section, the exchange contribution by SGGA is expected to be greater than the LSDA \cite{Ryee}. This tendency is clearly shown in Fig.~\ref{sro_dos}(b). In SGGA+$U$, the moment size is further enhanced ($M = 1.96$ $\mu_B$/f.u.) than LSDA+$U$ ($M = 1.67$ $\mu_B$/f.u.). On the other hand, in the case of CDFT+$U$ (Fig.~\ref{sro_dos}(a)), GGA+$U$ gives basically the same result with LDA+$U$ ($M = 1.41$ $\mu_B$/f.u.).

\subsection*{BaFe$_2$As$_2$} \label{Ba122}
The superconducting Fe pnictides have been a subject of intensive research activities. From the viewpoint of first-principles calculations, the unusually large magnetic moment by SDFT compared to experiments is a long standing issue \cite{Mazin,Yin,pnictides,Johannes}. Interestingly, to reproduce experimental moments, {\it negative} $U$ values within SDFT+$U$ \cite{Nakamura,Yi} have been adopted. As pointed out in Ref.~\cite{Yi}, however, it is hard to be justified in the physics sense. Here we note that the intrinsic exchange contribution of $\sim IM$ in SDFT can be too large as discussed in the above, and SDFT may not be the right starting point to take the correlation effects into account.

We found that CDFT+$U$ can provide much more sensible picture for magnetism in this material. Table~\ref{ba122} shows the calculated magnetic moment for BaFe$_2$As$_2$ with cRPA (constrained random phase approximation) value of $U=2.3$ eV \cite{Biermann_IBS}. The result of $M^\textrm{cFLL}$ is in a fairly good agreement with experiment ($M \simeq 0.9$ $\mu_B$/Fe \cite{Huang}) for $J = 0.3$ -- $0.5$ eV whereas $M^\textrm{sFLL}$ always overestimate the moments. Note that the reasonable size of $M$ is reproduced with realistic value of $U$ and $J$ only within CDFT+$U$. As shown in Table~\ref{ba122}, the moment is also sensitive to the way of defining local DM projector since the `full' projector tends to take the smaller on-site electron occupation compared to the `dual' \cite{MJH}.
The best comparison with experiment is achieved with $J=0.3$ eV for `dual' and $J \simeq 0.6$ eV for `full' projector. 

Also noticeable is the different $J$ dependence of moment by two functionals; $M^\textrm{cFLL}$ ($M^\textrm{sFLL}$) increases (decreases) as $J$ increases. This feature is again consistent with the behavior discussed in the previous section. The consistent result of cFLL with experiment is impressive even though the dynamic correlation beyond DFT+$U$ certainly plays the role in this system \cite{Georges_Hund}.

\section*{Summary and Conclusion}
We performed a comparative analysis on DFT+$U$ functionals employing two widely-used double counting forms and their relation to standard XC functionals. The detailed investigations on each formulation as well as the real material examples provided a clear understanding of different behaviors of DFT+$U$ functionals. The calculated energetics and spin potentials for representative model systems clearly show the role of double counting and XC functional in determining the ground state magnetic property. Competition between the effect of $J$ and the spin density XC energy is the key to understand the SDFT+$U$ result. Application to real materials including MnO, NiO, SrRuO$_3$, and BaFe$_2$As$_2$ further clarify the different tendency between the formalisms, supporting the analyses with model systems. As a rule of thumb, CDFT+$U$ is suggested as the desirable choice for most purposes.

\section*{Acknowledgements}
This research was supported by Basic Science Research Program through the National Research Foundation of Korea (NRF) funded by the Ministry of Education (2018R1A2B2005204). 
The computing resource was partly supported by
National Institute of Supercomputing and Networking / Korea Institute
of Science and Technology Information with supercomputing resources
including technical support (KSC-2015-C2-011).

\section*{Author contributions}

S.R. performed the calculations and analysis under the supervision of M.J.H. Both authors wrote the manuscript. 

\section*{Competing interests}

The authors declare no competing financial interests.

\bibliography{ref}

\begin{thebibliography}{10}
\bibitem{sum_rule}
\bibinfo{note}{The double counting form of AMF formalism is simplified by using
  the sum rule of Coulomb interaction matrix (Ref.~\cite{Czyzyk});
  $\sum_{m}\langle m m'|V_{ee}|m m' \rangle = (2l+1)U$ and $\sum_{m}\langle m
  m'|V_{ee}|m' m \rangle = U+2lJ$.}
\bibitem{comment}
\bibinfo{note}{We could not find the direct estimation of $J$ for the
  $d$-shell. Considering the exchange coupling within $t_{2g}$ orbitals of
  Slater-Kanamori Hamiltonian (Ref.~\cite{Si,Dang}), $J$ is estimated to be
  $\sim 0.4$ -- $0.6$ eV.}
\end{thebibliography}


\newpage

\newpage

\newpage

\newpage

\newpage

\newpage
\begin{table} [!htbp]
	\renewcommand{\arraystretch}{1.2}
	\begin{tabular}{ l | c c l c }
		\hline \hline
		Configuration &\ \ $M$ &\ \ $N$ &\ \  Occupation  &\ \   $\alpha$ 	\\
		\hline		
		\rule{22pt}{0ex} 1	&\ \ 1 &\ \ 1  	&\ \  $|00100;00000 \rangle$  &\ \	 $d_{xy}$ 	 \\
		
		\rule{22pt}{0ex} 1'  &\ \ 1  &\ \ 1    &\ \ $|00\frac{1}{3}\frac{1}{3}\frac{1}{3};00000 \rangle$  &\ \     $d_{xy},d_{zx},d_{yz}$ 	 \\
		
		\rule{22pt}{0ex} 2   &\ \ 1  &\ \ 5    &\ \ $|00111;00011 \rangle$  &\ \     $d_{xy}$ 	\\
		
		\rule{22pt}{0ex} 2'   &\ \ 1  &\ \ 5    &\ \ $|00111;00\frac{2}{3}\frac{2}{3}\frac{2}{3} \rangle$  &\ \     $d_{xy},d_{zx},d_{yz}$ 	\\
		
		\rule{22pt}{0ex} 3   &\ \ 1  &\ \ 7    &\ \ $|10111;00111 \rangle$  &\ \     $d_{z^2}$ 	\\
		
		\rule{22pt}{0ex} 3'   &\ \ 1  &\ \ 7    &\ \ $|\frac{1}{2}\frac{1}{2}111;00111 \rangle$  &\ \     $d_{z^2},d_{x^2-y^2}$ 	\\
		
		\rule{22pt}{0ex} 4   &\ \ 1  &\ \ 9    &\ \ $|11111;01111 \rangle$  &\ \     $d_{z^2}$ 	\\
		
		\rule{22pt}{0ex} 4'   &\ \ 1  &\ \ 9    &\ \ $|11111;\frac{1}{2}\frac{1}{2}111 \rangle$  &\ \     $d_{z^2},d_{x^2-y^2}$ 	\\
		
		\rule{22pt}{0ex} 5   &\ \ 2  &\ \ 2    &\ \ $|00110;00000 \rangle$  &\ \     $d_{xy},d_{zx}$ 	\\
		
		\rule{22pt}{0ex} 5'   &\ \ 2  &\ \ 2    &\ \ $|00\frac{2}{3}\frac{2}{3}\frac{2}{3};00000 \rangle$  &\ \     $d_{xy},d_{zx},d_{yz}$ 	\\
		
		\rule{22pt}{0ex} 6   &\ \ 2  &\ \ 4    &\ \ $|00111;00100 \rangle$  &\ \     $d_{zx},d_{yz}$ 	\\
		
		\rule{22pt}{0ex} 6'   &\ \ 2  &\ \ 4    &\ \ $|00111;00\frac{1}{3}\frac{1}{3}\frac{1}{3} \rangle$  &\ \     $d_{xy},d_{zx},d_{yz}$ 	\\
		
		\rule{22pt}{0ex} 7   &\ \ 2  &\ \ 8    &\ \ $|11111;00111 \rangle$  &\ \     $d_{z^2},d_{x^2-y^2}$ 	\\
		
		\rule{22pt}{0ex} 8   &\ \ 3  &\ \ 3    &\ \ $|00111;00000 \rangle$  &\ \     $d_{xy},d_{zx},d_{yz}$ 	\\
		
		\rule{22pt}{0ex} 9   &\ \ 3  &\ \ 7    &\ \ $|11111;00011 \rangle$  &\ \     $d_{xy}$ 	\\
		
		\rule{22pt}{0ex} 9'   &\ \ 3  &\ \ 7    &\ \ $|11111;00\frac{2}{3}\frac{2}{3}\frac{2}{3} \rangle$  &\ \     $d_{xy},d_{zx},d_{yz}$ 	\\
		
		\rule{22pt}{0ex} 10   &\ \ 4  &\ \ 4    &\ \ $|01111;00000 \rangle$  &\ \     $d_{xy}$ 	\\
		
		\rule{22pt}{0ex} 10'   &\ \ 4  &\ \ 4    &\ \ $|\frac{1}{2}\frac{1}{2}111;00000 \rangle$  &\ \     $d_{xy},d_{zx},d_{yz}$ 	\\
		
		\rule{22pt}{0ex} 11   &\ \ 4  &\ \ 6    &\ \ $|11111;00100 \rangle$  &\ \     $d_{zx},d_{yz}$ 	\\
		
		\rule{22pt}{0ex} 11'   &\ \ 4  &\ \ 6    &\ \ $|11111;00\frac{1}{3}\frac{1}{3}\frac{1}{3} \rangle$  &\ \     $d_{xy},d_{zx},d_{yz}$ 	\\
		
		\rule{22pt}{0ex} 12   &\ \ 5  &\ \ 5    &\ \ $|11111;00000 \rangle$  &\ \     $d_{xy},d_{zx},d_{yz}$ 	\\
		\hline \hline
	\end{tabular}
	\caption{The electronic configurations considered in Fig.~\ref{potential}. In the fourth column, $d$-shell occupations are presented in a form of $|n^{\uparrow}_{z^2} n^{\uparrow}_{x^2-y^2} n^{\uparrow}_{xy} n^{\uparrow}_{zx} n^{\uparrow}_{yz}; n^{\downarrow}_{z^2} n^{\downarrow}_{x^2-y^2} n^{\downarrow}_{xy} n^{\downarrow}_{zx} n^{\downarrow}_{yz}  \rangle$ where $n^{\sigma}_{m}$ denotes the number of electrons occupied in the $m$ orbital with spin $\sigma$. The primed configurations refer to the fractional occupations. The magnetic moment $M$ (in the unit of $\mu_B$) and the number of electrons $N$ are given in the second and third column, respectively. The $\alpha$ are chosen to represent the lowest unoccupied or partially occupied down-spin orbitals assuming octahedral environment. }
	\label{conf}
\end{table}

\newpage
\begin{table} [!htbp] 
	\renewcommand{\arraystretch}{1.3}
	\begin{tabular}{ c | c | c c c }
		\hline \hline
		$U$ [eV] &\ \ $J$ [eV] &\ \ DM &\ \  $M^\textrm{cFLL}$ [$\mu_B$/Fe]  &\ \ $M^\textrm{sFLL}$ [$\mu_B$/Fe] 	\\
		\hline \hline	
		2.3	&\ \ 0.3  &\ \ dual  	&\ \  0.94  &\ \ 2.82	 \\
		&\ \      &\ \ full  	&\ \  0.29  &\ \ 2.63	 \\
		
		&\ \ 0.5  &\ \ dual     &\ \  1.78	&\ \ 2.77	 \\
		&\ \      &\ \ full     &\ \  0.75	&\ \ 2.59 	 \\
		
		&\ \ 0.7  &\ \ dual     &\ \  2.34	&\ \ 2.73	 \\
		&\ \      &\ \ full     &\ \  1.33	&\ \ 2.56	 \\
		\hline \hline
		
	\end{tabular}
	\caption{Calculated magnetic moment of BaFe$_2$As$_2$ by cFLL ($M^\textrm{cFLL}$) and sFLL ($M^\textrm{sFLL}$). Experimental crystal structure in spin-density-wave phase is used \cite{Huang}. Two different definitions of DM are used for the comparison (namely, `dual' and `full'). }
	\label{ba122}
\end{table}


\newpage

\end{document}